\documentclass{article}
\title{On the Closed Form Solution for the Geodesics in SdS Space}
\author{Richard J. Drociuk\footnote{drociuk@sfu.ca}}
\begin{document}
\maketitle
\begin{center}
``Dedicated to Niels Henrik Abel; The Great Pioneer into the Theory of Functions and Algebraic Equations"
\end{center}
\section*{}
The closed form solution for the geodesics of classical particles in SdS space is 
obtained in terms of hyperelliptic modular 
functions and multiple hypergeometric functions. The closed form solution for the five 
roots of the fifth degree polynomial is found giving the branch places on the genus 
two Riemann surface. `The Inversion Problem', for the genus two hyperelliptic integral, is solved in a closed form.  The solution 
is shown to reduce to elliptic functions when the        
cosmological constant is zero. 
Current observational data is in favor of a cosmological constant model${}^{1}$. This 
solution is important in astrophysical 
applications of measuring the cosmological constant.
\section*{\uppercase\expandafter{\romannumeral1}. Introduction} 
To calculate the geodesics in the centrally symmetric gravitational 
field of a Schwarz\-schild black hole in a universe with a cosmological 
constant($\Lambda\neq 0$), i.e. Schwarz\-schild de-Sitter space(SdS space), the 
Hamilton-Jacobi method is applied to the 
metric${}^{2-5}$
\begin{equation}
ds^2  = g_{tt}dt^2+g_{rr}dr^2+g_{\phi\phi}d\phi^2+g_{\theta\theta}d\theta^2
\end{equation} 
\begin{equation}
g_{tt}  =  (1-\frac{1}{3}\sigma\Lambda r^2-\frac{r_{g}}{r})c^2,
g_{rr} =  -(1-\frac{1}{3}\sigma\Lambda r^2-\frac{r_{g}}{r})^{-1},
g_{\phi\phi} = -r^2 sin^2\theta,
g_{\theta\theta}  =  -r^2
\end{equation}
Where $\sigma=\pm 1$ depending on whethor or not the universe is 
de-Sitter or Anti-de-Sitter, respectively. $\Lambda$ is the cosmological constant 
and was first introduced by Einstein${}^{6}$ and $r_{g}$ was introduced by Schwarzschild. 
The functional form of the geodesics when $r_{g}=0$ have been studied, they are 
given in terms of genus zero circular functions. When $\Lambda=0$ the geodesics are 
given by genus one elliptic functions${}^{7}$. In this paper , the functional form of the 
geodesics when $\Lambda$ and $r_{g}$ are both non-zero are found, in terms of genus 
two hyperelliptic theta functions${}^{8-9}$and multiple hypergeometric functions${}^{10}$.
\section*{\uppercase\expandafter{\romannumeral2}. Derivation of the Hyperelliptic 
$\phi$-Integral}
Let a particle of mass $m$ be constrained to move in the equatorial plane, 
$\theta=0$, and let the action be given by${}^{11}$,
\begin{equation}
S = -Et+L\phi+S_{r}
\end{equation}
where $E$ is the total energy of the particle and $L$ is it's angular momentum. By 
the conservation of four momentum,
\begin{equation}
g^{ij}\frac{dS}{dx^i}\frac{dS}{dx^j}=m^2 c^2
\end{equation}
we can insert $S$ from (3) and $g^{ij}$ from (2) into (4) and obtain,
\begin{equation}
S_{r} = \int_{r_{0}}^{r} \frac{(\frac{E^2}{c^2}-(1-\frac{1}{3} \Lambda \sigma 
r^2-\frac{r_{g}}{r})(\frac{L^2}{r^2}+m^2c^2))^{\frac{1}{2}}}{(1-\frac{1}{3}\Lambda\sigma 
r^2-\frac{r_{g}}{r})}.
\end{equation}
Differentiating $S_{r}$ w.r.t. $L$ and using
\begin{equation}
\phi+\frac{\partial S_{r}}{\partial L}=const=0
\end{equation}
\begin{equation}
\phi =  \int_{r_{0}}^{r} \frac{L dr}{r^2(\frac{E^2}{c^2}-(1-\frac{1}{3} \Lambda 
\sigma
r^2-\frac{r_{g}}{r})(\frac{L^2}{r^2}+m^2c^2))^{\frac{1}{2}}}
\end{equation}
This integral is a hyperelliptic integral of genus, $g=2$. The problem is to invert 
this integral, i.e. express the upper limit $r$ as a function of 
$\phi$,${}^{12-14}$ and $\phi$ as a function of $r$, this is known as the `Inversion Problem'. The inversion problem was first 
conceived by Euler${}^{15}$. 
\section*{\uppercase\expandafter{\romannumeral3}. The Solution to the Inversion 
Problem for the Hyperelliptic Integral of Arbitrary Genus}
The following will be an account given by H.F. Baker${}^{8-9}$ and H. Exton${}^{10}$ on the 
application of theta functions
and multiple hypergeometric functions to obtain the closed form solution 
to the 
arbitrary genus hyperelliptic integral, except for the branch places on the Riemann 
surface. This requires the solution to the $n^{th}$ degree polynomial, for its n 
roots. 
The general solution to the inversion 
problem is given by solving $g$ of the $2g+1$ equations, for the $g$ variable places 
$x_{1},\ldots,x_{g}$,
\begin{equation}
\frac{\vartheta^2(u|u^{b,a})}{\vartheta^2(u|u^{b',a'})}=A(b)(b-x_{1})\ldots(b-x_{g})
\end{equation}
The notation Baker uses for the generalized theta function is,
\begin{equation}
\vartheta(u|u_{b,a})=\vartheta(u|\frac{1}{2}\Omega_{m,m'})=\vartheta(u;\frac{1}{2}m,\frac{1}{2}m')=e^{au^2}\Theta(v;\frac{1}{2}m,\frac{1}{2}m')
\end{equation}
where $a$ is an arbitrary $g\times g$ symmetrical matrix, since putting (9) into (8) gives,
\begin{equation}
\frac{\vartheta^2(u|u^{b,a})}{\vartheta^2(u|u^{b',a'})}=\frac{\Theta^2(v|\frac{1}{2}m,\frac{1}{2}m')}{\Theta^2(v|\frac{1}{2}k,\frac{1}{2}k')}
\end{equation}
where $m$, $m'$, $k$ and $k'$ are integers. Riemann's theta functions are defined as,
\begin{equation}
\Theta(v;\frac{1}{2}m,\frac{1}{2}m')=\Sigma e^{2hv(n+\frac{1}{2}m')+b(n+\frac{1}{2}m')^2+i \pi m(n+\frac{1}{2}m')}
\end{equation}
with, $\Sigma=\sum_{n_{1}=-\infty}^{+\infty} \ldots \sum_{n_{g}=-\infty}^{+\infty}$. $h$ is a $g \times g$ matrix, in general 
asymmetrical. $b$ is a $g \times g$ symmetrical matrix. In (10), u denotes the 
$g$ quantities,
\begin{equation}
u_{i}^{x_{1},a_{1}}+\ldots+u_{i}^{x_{g},a_{g}}=u_{i}
\end{equation}
with $i=1,\ldots,g$. This is known as Abel's Theorem. Riemann's Normal Integral of the first kind is defined as,
\begin{equation}
u_{i}^{x,a}= \int_{a}^{x} \frac{(x,1)_{i,g-1}dx}{y}
\end{equation}
where
\begin{equation}
(x,1)_{i,g-1} = A_{i,g-1} x^{g-1}+ \ldots +A_{i,0}
\end{equation}
and the Abel coefficients $A_{i,g-1},\ldots,A_{i,0}$ are constants determined experimentally. The denominator in (13) is 
given by,
\begin{equation}
y^2 = 4(x-a_{1})\ldots(x-a_{g})(x-c_{1})\ldots(x-c_{g})(x-c)
\end{equation}
where $a_{1},\ldots,a_{g},c_{1},\ldots,c$ are branch places and $x$ is a variable on the one complex dimensional Riemann surface of 
genus, $g$. Now since the sum of terms is finite in (13), then by combining (13)(14)and(15) an arbitrary term in $u_{i}^{x,a}$ may be 
transformed to Extonian form${}^{10}$, using a M\"obius transformation,
\begin{eqnarray}
  \int_{0}^{z} 
t^{(a-1)}(1-t)^{(c-a-1)}(1-tz_{1})^{(-\frac{1}{2})}\ldots(1-tz_{2g})^{(-\frac{1}{2})}dt{}\nonumber\\
  {} =    \frac{z^a}{a} 
F_{D}^{(2g+1)}(a,\frac{1}{2},\ldots,\frac{1}{2},1+a-c;a+1;z_{1},\ldots,z_{2g},z)
\end{eqnarray}
where $Re(a)$ and $Re(c-a)$ are greater than zero. $a$ and $c$ 
are appropriately chosen constants. The variables $z_{1},\ldots, 
z_{n}$ depend on the roots of $y$ and the lower limit. While $z$ depends on the upper 
limit of the hyperelliptic integral. The condition that $z_{1},\ldots,z_{2g}$ and $z$ 
each have their absolute value less than one ensures that the Lauricella function 
$F_{D}^{(2g+1)}$ converges, making the variables in Abel's theorem well defined. The 
convergence takes place in a unit sphere in $2g+1$ hyperspace. The 
Lauricella function has the expansion,
\begin{eqnarray}
F_{D}^{(2g+1)} 
(a,\frac{1}{2},\ldots,\frac{1}{2},1+a-c;a+1;z_{1},\ldots,z_{2g},z){}\nonumber\\
{}=\sum_{(m)=-\infty}^{+\infty} 
\frac{(a,m_{1}+\ldots+m_{2g+1})(\frac{1}{2},m_{1})\ldots(1+a-c,m_{2g+1})z_{1}^{m_{1}}\ldots 
z^{m_{2g+1}}}{(a+1,m_{1}+\ldots+m_{2g+1}){m_{1}!}\ldots {m_{2g+1}!}}
\end{eqnarray}
\begin{equation}
|z_{1}|<1 ,\ldots , |z_{2g}|<1 , |z|<1
\end{equation}
where $(m) = m_{1}\ldots m_{2g}$ and $z_{i}$ have the form from the M\"obius Transformation,
\begin{equation}
z_{i}=\frac{r_{j}-a}{r_{i}-a}
\end{equation}
where $a$ is the lower limit of the hyperelliptic integral(13), and
\begin{equation}
z = \frac{x-a}{r_{i}-a}
\end{equation}
Now let,
\begin{equation}
\frac{1}{2} \Omega_{m,m'} = 
m_{1}\omega_{r,1}+\ldots+m_{g}\omega_{r,g}+m'_{1}\omega'_{r,1}+\ldots+m'_{g}\omega'_{r,g}
\end{equation}
where the $\omega$ have the representation,
\begin{equation}
\omega = C \times 
F_{D}^{(2g-1)}(a,\frac{1}{2},\ldots,\frac{1}{2};c;z_{1},\ldots,z_{2g-1})
\end{equation}
and where $m$ and $m'$ are equal to $+1$ when integrating across a period loop right to left, and 
$-1$ when crossing from left to right and $0$ when not crossing a period loop. 
${C}$ is a constant that depends on the 
branch places and Abel's coefficients. $\omega$ is convergent in a unit sphere in $2g-1$ hyperspace. $a$ and $c$ are the 
appropriate constants. Now define 
an integral of the first kind via,
\begin{equation}
\pi i v_{r}^{x,a}=h_{r,1} u_{1}^{x,a}+\ldots+h_{r,g}u_{g}^{x,a}
\end{equation}
with $r=1,\ldots,g$ and where,
\begin{equation}
2 h \omega = \pi i
\end{equation}
\begin{equation}
2 h \omega' = \pi i \tau
\end{equation}
where Riemann's condition for hyperelliptic integrals is imposed${}^{16-18}$. 
\begin{equation}
\tau = \tau^t   
\end{equation}
With the above equations, Riemann's theta function becomes,
\begin{equation}
\Theta(v;\frac{1}{2}m,\frac{1}{2}m')= \Sigma e^{2 \pi i v (n+\frac{1}{2}m')+i \pi \tau 
(n+\frac{1}{2}m')^2 + m (n+\frac{1}{2}m')}
\end{equation}
where the quadratic forms are given by,
\begin{equation}
v(n+\frac{1}{2}m')=vn+\frac{1}{2}m'v=v_{1}n_{1}+\ldots+v_{g}n_{g}+\frac{1}{2}v_{1}m'_{1}+\ldots+\frac{1}{2}v_{g}m'_{g}
\end{equation}
and similarly,
\begin{eqnarray}
\tau (n+\frac{1}{2}m') & = & \tau n^2 + \tau n m' + \frac{1}{4} \tau m'^2{}\nonumber\\
 & & {}= (\tau_{1,1}n_{1}^2+2\tau_{1,2}n_{1}n_{2}+\ldots+\tau_{g,g}{n_{g}}^2){}\nonumber\\
 & & {}+\sum_{s=1}^{g} \sum_{r=1}^{g} \tau_{r,s} n_{r} m'_{s}{}\nonumber\\
 & & {}+\frac{1}{4}(\tau_{1,1} {m'_{1}}^2+2 \tau_{1,2} m'_{1} m'_{2}+\ldots+\tau_{g,g} {m'_{g}}^2)
\end{eqnarray}
It has been proven by Riemann that for the integral of the first kind, when it has the given period scheme, then the 
imaginary part of,
\begin{equation}
\tau_{1,1} {n_{1}}^2+2 \tau_{1,2} n_{1} n_{2}+\ldots+\tau_{g,g} {n_{g}}^2
\end{equation}
is positive for all integer values of $n_{1}$ and $n_{2}$, with the exception of $n_{1}=n_{2}=0$. Therefore the modulus of 
$e^{i \pi \tau n^2}$ is less than unity and the function(27) converges for all values of its argument, $v$. The last thing to 
define in (8), is the constant $A(b)$, it is defined as,
\begin{equation}
A(b) = (\epsilon \frac{d}{dx} (x-a_{1})\ldots(x-a_{g})(x-c_{1})\ldots(x-c_{g})(x-c)|_{x=b})^{-\frac{1}{2}}
\end{equation}
where $\epsilon$ is $+1$ when $u^{b,a}$ is an odd half period and $-1$ when it is an even half period.
\section*{\uppercase\expandafter{\romannumeral4}. The Complete Solution to the Inversion Problem of the Genus Two Hyperelliptic 
Integral}
The solution to the inversion of the genus two hyperelliptic integral consists of two parts. The first part is to express the 
upper 
limits, $x_{1}$ and $x_{2}$, in terms of the angles, $u_{1}$ and $u_{2}$, in Abel's theorem,
\begin{equation}
{u_{1}}^{x_{1},a_{1}}+{u_{1}}^{x_{2},a_{2}}=u_{1}
\end{equation}
\begin{equation}
{u_{2}}^{x_{1},a_{1}}+{u_{2}}^{x_{2},a_{2}}=u_{2}
\end{equation}
and secondly, express the angles  $u_{1}$ and $u_{2}$ in terms of $x_{1}$ and $x_{2}$, so that $u_{1}$, $u_{2}$, 
$x_{1}$ and 
$x_{2}$ become well defined coordinates on Kummer's quartic surface${}^{19}$.
The solution to the first part is given by${}^{20-22,8-9}$, with $g=2$,
\begin{equation}
\frac{\Theta^2(v_{1},v_{2};\frac{1}{2}m_{1},\frac{1}{2}m_{2},\frac{1}{2}m'_{1},\frac{1}{2}m'_{2})}{\Theta^2(v_{1},v_{2};\frac{1}{2}k_{1},\frac{1}{2}k_{2},\frac{1}{2}k'_{1},\frac{1}{2}k'_{2})}=A(b)(b-x_{1})(b-x_{2})
\end{equation}
where by (23) we have,
\begin{equation}
\pi i v_{1} =\pi i v_{1}^{x,a}=h_{1,1} u_{1}^{x,a}+h_{1,2}u_{2}^{x,a}
\end{equation}  
\begin{equation}
\pi i v_{2} =\pi i v_{2}^{x,a}=h_{2,1} u_{1}^{x,a}+h_{2,2}u_{2}^{x,a}
\end{equation}
where the $h_{i,j}$ are the elements of the following matrix,
\begin{displaymath}
\mathbf{h}= \frac{\pi i}{ \Delta}
\left( \begin{array}{ccc}
\omega_{2,2}&-\omega_{1,2}&\\
-\omega_{2,1}&\omega_{1,1}&\\
\end{array} \right)
\end{displaymath}
and
\begin{equation}
\Delta = det(\omega) \neq 0
\end{equation}
Now the solution to the second part of the inversion problem is facilitated by first choosing,
\begin{equation}
x_{2}=a_{2}
\end{equation}
by (13), we have
\begin{equation}
u_{1}=u_{1}^{x_{1},a_{1}}= \int_{a_{1}}^{x_{1}} \frac{(A_{1,0}+A_{1,1}x)dx}{y}
\end{equation} 
\begin{equation}
u_{2}=u_{2}^{x_{1},a_{1}}= \int_{a_{1}}^{x_{1}} \frac{(A_{2,0}+A_{2,1}x)dx}{y}
\end{equation}
where,
\begin{equation}
y^2 = 4x^5+\lambda_{4}x^4+\lambda_{3}x^{3}+\lambda_{2}x^2+\lambda_{1}x+\lambda_{0}
\end{equation}
which can be rewritten as,
\begin{equation}
y^2=4(x-a_{1})(x-a_{2})(x-c_{1})(x-c_{2})(x-c)
\end{equation}
Now let (39) and (40) be written as,
\begin{equation}
{u_{k}}^{x,y}=\int_{y}^{x} 
\frac{(A_{k,0}+A_{k,1}t)dt}{\sqrt{4(t-r_{i_{1}})(t-r_{i_{2}})(t-r_{i_{3}})(t-r_{i_{4}})(t-r_{i_{5}})}}
\end{equation}
where $k=1,2$. $i_{1},i_{2},i_{3},i_{4}$ and $i_{5} \in {\aleph \hspace{0.1cm}}  mod  {\hspace{0.1cm} 5}$ where $\aleph 
\in [0,1,\ldots \infty)$, so that,
\begin{equation}
r_{0},r_{1},r_{2},r_{3},r_{4} \in (a_{1},a_{2},c_{2},c_{1},c)
\end{equation}
The integral(43) may written as a product of factors, via the M\"obius 
transformation,
\begin{equation}
z=\frac{x-y}{r_{i_{1}}-y}
\end{equation}
we have,
\begin{eqnarray}
{u_{k}}^{x,y} & = & B(A_{k,0}+A_{k,1}y)\times{}\nonumber\\ 
 & & {} 
\int_{0}^{z}(1-z)^{-\frac{1}{2}}(1-zx_{i_{2}})^{-\frac{1}{2}}(1-zx_{i_{3}})^{-\frac{1}{2}}(1-zx_{i_{4}})^{-\frac{1}{2}}(1-zx_{i_{5}})^{-\frac{1}{2}}dz 
{}
\nonumber\\
 & & {}+B(r_{i_{1}}-y)A_{k,1} \times{}\nonumber\\
 & & {} 
\int_{0}^{z}z(1-z)^{-\frac{1}{2}}(1-zx_{i_{2}})^{-\frac{1}{2}}(1-zx_{i_{3}})^{-\frac{1}{2}}(1-zx_{i_{4}})^{-\frac{1}{2}}(1-zx_{i_{5}})^{-\frac{1}{2}}dz\nonumber\\
\end{eqnarray}
where
\begin{equation}
B \equiv B(y,r_{i_{1}},r_{i_{2}},r_{i_{3}},r_{i_{4}},r_{i_{5}})= \frac{1}{2i} 
\sqrt{\frac{r_{i_{1}}-y}{(r_{i_{2}}-y)(r_{i_{3}}-y)(r_{i_{4}}-y)(r_{i_{5}}-y)}}
\end{equation}
and
\begin{equation}
x_{i_{2}}=\frac{r_{i_{1}}-y}{r_{i_{2}}-y};x_{i_{3}}=\frac{r_{i_{1}}-y}{r_{i_{3}}-y};x_{i_{4}}=\frac{r_{i_{1}}-y}{r_{i_{4}}-y};x_{i_{5}}=\frac{r_{i_{1}}-y}{r_{i_{5}}-y}
\end{equation}
The two integrals(46) are of Extonian form, so that the integral(43)is given by,
\begin{equation}
{u_{k}}^{x,y}=phef(A_{k,0},A_{k,1};x,y)+shef(A_{k,1};x,y)
\end{equation}
where ``phef" stands for ``primary hyper-elliptic function" and ``shef" stands for ``secondary hyper-elliptic function". They are 
defined 
as,
\begin{equation}
phef(A_{k,0},A_{k,1};x,y)\equiv 
 B(A_{k,0}+A_{k,1}y)z{F_{D}}^{(5)}(1,\frac{1}{2},\frac{1}{2},\frac{1}{2},\frac{1}{2},\frac{1}{2};2;x_{i_{2}},x_{i_{3}},x_{i_{4}},x_{i_{5}},z)
\end{equation}
\begin{equation}
shef(A_{k,1};x,y)\equiv
\frac{1}{2}B 
A_{k,1}(r_{i_{1}}-y)z^2{F_{D}}^{(5)}(2,\frac{1}{2},\frac{1}{2},\frac{1}{2},\frac{1}{2},\frac{1}{2};3;x_{i_{2}},x_{i_{3}},x_{i_{4}},x_{i_{5}},z)
\end{equation}
Both (50) and (51) are defined via the twelve slotted, five variable, Lauricella D multiple hypergeometric function, 
${F_{D}}^{(2g+1)}$, with $2g+1=5$ Ref. 10. 
These functions converge when the moduli of the five variable parameters $x_{i_{2}}$, 
$x_{i_{3}}$, $x_{i_{4}}$, $x_{i_{5}}$ and $z$ have 
their absolute values less than unity. The covergence takes place in a unit sphere in five dimensional hyperspace. In 
this case (50) and (51) may be expanded into their Lauricella series, i.e.
\begin{eqnarray}
& &phef(A_{k,0},A_{k,1};x,y) =  B(A_{k,0}+A_{k,1}y)\times {}\nonumber\\ 
& 
&{}\sum_{(m)=0}^{+\infty}\frac{(1,m_{1}+m_{2}+m_{3}+m_{4}+m_{5})(\frac{1}{2},m_{1})(\frac{1}{2},m_{2})(\frac{1}{2},m_{3})(\frac{1}{2},m_{4})(\frac{1}{2},m_{5})}{(2,m_{1}+m_{2}+m_{3}+m_{4}+m_{5})m_{1}!m_{2}!m_{3}!m_{4}!m_{5}!}{}\nonumber\\
& & {} \times {x_{i_{2}}}^{m_{1}}{x_{i_{3}}}^{m_{2}}{x_{i_{4}}}^{m_{3}}{x_{i_{5}}}^{m_{4}}z^{m_{5}+1}
\end{eqnarray}
\begin{eqnarray}
& &shef(A_{k,1};x,y) = \frac{1}{2} BA_{k,1}(r_{i_{1}}-y)\times {}\nonumber\\
&
&{}\sum_{(m)=0}^{+\infty}\frac{(2,m_{1}+m_{2}+m_{3}+m_{4}+m_{5})(\frac{1}{2},m_{1})(\frac{1}{2},m_{2})(\frac{1}{2},m_{3})(\frac{1}{2},m_{4})(\frac{1}{2},m_{5})}{(3,m_{1}+m_{2}+m_{3}+m_{4}+m_{5})m_{1}!m_{2}!m_{3}!m_{4}!m_{5}!}{}\nonumber\\
& & {} \times {x_{i_{2}}}^{m_{1}}{x_{i_{3}}}^{m_{2}}{x_{i_{4}}}^{m_{3}}{x_{i_{5}}}^{m_{4}}z^{m_{5}+2}
\end{eqnarray}
with
\begin{equation}
|x_{i_{2}}|<1;|x_{i_{3}}|<1;|x_{i_{4}}|<1;|x_{i_{5}}|<1;|z|<1
\end{equation}
The quadrupoly periodic double theta function is given from (27) to (29),
\begin{eqnarray}
\Theta(v_{1},v_{2};\frac{1}{2}m_{1},\frac{1}{2}m_{2},\frac{1}{2}{m'}_{1},\frac{1}{2}{m'}_{2})
= \sum_{n_{1}=-\infty}^{\infty} \sum_{n_{2}=-\infty}^{\infty} e^{\pi i(v_{1}(2n_{1}+{m'}_{1})+v_{2}(2n_{2}+{m'}_{2}))}\nonumber\\
\times  q^{(2n_{2}+{m'}_{2})^2}
     {q'}^{(2n_{1}+{m'}_{1})^2}    r^{(2n_{1}+{m'}_{1})(2n_{2}+{m'}_{2})}
(-1)^{(m_{1}n_{1}+m_{2}n_{2})}      i^{(m_{1}{m'}_{1}-m_{2}{m'}_{2})}\nonumber\\
\end{eqnarray}
where
\begin{equation}
q = e^{\frac{1}{4} i \pi \tau_{1,1}},
q'  =  e^{\frac{1}{4} i \pi \tau_{2,2}},
r  =  e^{\frac{1}{2} i \pi \tau_{1,2}}
\end{equation}  
\begin{equation}
m_{1} = \rho = \rho,
m_{2}  =  {\rho}' = \lambda,
{m'}_{1}  =  \sigma = \nu,
{m'}_{2}  =  {\sigma}'= \mu
\end{equation}  
so that $a_{r}$ in Forsyth's paper is equal to
\begin{equation}
a_{r} = (i)^{(\sigma' \rho'+\rho \sigma)} r^{(2n+\sigma')(2m+\sigma)}
\end{equation}
where $n=n_{1}$ and $m=n_{2}$ are the summation indices. Forsyth gives a review of
Rosenhain's theory,${}^{22-23}$ of the fifteen ratios of the quadrupoly periodic theta
functions. From the list given, the first ratio is selected and is equal to,
\begin{equation}
\frac{\Theta^2(v_{1},v_{2};\frac{1}{2},\frac{1}{2},0,\frac{1}{2})}{\Theta^2(v_{1},v_{2};\frac{1}{2},\frac{1}{2},0,0)}=
\sqrt{\frac{a_{1}-a_{2}}{(a_{1}-c_{1})(a_{1}-c_{2})(a_{1}-c)}} (x_{1}-c)
\end{equation}
where (31)(38)(55)and (59) have been used. The double theta function used in (59) is,
\begin{eqnarray}
\theta_{13}= \sum_{n_{1}=-\infty}^{+\infty} \sum_{n_{2}=-\infty}^{+\infty} e^{\pi i (v_{1} (
2n_{1})+v_{2}(2n_{2}+1))}{}\nonumber\\
{}  \times q^{(2n_{2}+1)^2}  {q'}^{(2n_{1})^2}  r^{(2n_{1})(2n_{2}+1)}  (-1)^{(n_{1}+n_{2})} i^{(-1)}
\end{eqnarray}
where Forsyth's notation is adopted,
\begin{equation}
\theta_{13} = \Theta(v_{1},v_{2};\frac{1}{2},\frac{1}{2},0,\frac{1}{2})
\end{equation}
and in the denominator of (59), is
\begin{equation}
\theta_{12}
= \sum_{n_{1}=-\infty}^{+\infty} \sum_{n_{2}=-\infty}^{+\infty} e^{\pi i
(v_{1}2n_{1}+v_{2}(2n_{2}))}{q}^{(2n_{2})^2}{q'}^{(2n_1)^2}r^{(2n_{1})(2n_{2})}(-1)^{n_{1}+n_{2}}
\end{equation}
with,
\begin{equation}
\theta_{12} = \Theta(v_{1},v_{2};\frac{1}{2},\frac{1}{2},0,0).
\end{equation}  
This completes the inversion of the genus two hyperelliptic integral. The analytic continuation formulae are given in Exton. If 
one
wants the solution with all four coordinates $x_{1}$, $x_{2}$, $u_{1}$ and $u_{2}$ in (32) and (33) one does not choose 
$x_{2}=a_{2}$. Proceed as above adding on the appropriate multiple hypergeometric functions and selecting another theta function. This 
was not done here to keep the solution as concise as possible while keeping its closed form (although all four coordinates 
may 
play a 
role in the geometric interpretation of the solution when applied to the problem at hand). Now in order to have a closed form 
solution we 
must find the roots of the quintic, $r_{1}$, $ r_{2}$, $ r_{3}$, $ r_{4}$ and $r_{5}$ in a closed form.
\section*{\uppercase\expandafter{\romannumeral5}. The Complete Solution to the Quintic}
\subsection*{A. The Tschirnhausian Transformation to Bring's Normal Form}
The initial Tschirnhausian transformation  used is a generalization of 
Bring's${}^{24-25}$,
but a simplification of Cayley's${}^{26}$, with a quartic substitution,
\begin{equation} 
Eq1 = x^4+dx^3+cx^2+bx+a+y
\end{equation}
to the general quintic equation,
\begin{equation}
Eq2 = x^5+mx^4+nx^3+px^2+qx+r.
\end{equation}
Then by the process of elimination between (64) and (65), the
following 25  
equations are obtained,   
\begin{displaymath}
M_{15} = 1, M_{14} = d, M_{13} = c, M_{12} = b, M_{11} = a+y,M_{25} = m-d
\end{displaymath}
\begin{displaymath}    
M_{24} = n-c, M_{23} = p-b, M_{22} = -y+q-a, M_{21} = r, M_{35} = n+dm-m^2-c
\end{displaymath}
\begin{displaymath}
M_{34} = p-b-mn+dn, M_{33} = q-a-mp-y+dp, M_{32} = r-mq+dq, M_{31} = dr-mr
\end{displaymath}
\begin{displaymath}
M_{45} = -cm-m^3+b-p+dm^2+2mn-dn, M_{44} = a+dmn+y-dp-m^2n-q+n^2+mp-cn
\end{displaymath}
\begin{displaymath}
M_{43} = np-r+dmp-dq+mp-m^2p-cp, M_{42} = np-r+dr-m^2q-cq+dmq+nq,
\end{displaymath}
\begin{eqnarray}
M_{41}  & = & -m^2r-cr+nr+dmr\nonumber\\
 M_{55} & = & bm-2mp+q-y+cn-2dmn-a+3m^2n-cm^2\nonumber\\
 & & {}+dm^3-n^2-m^4+dp\nonumber\\
M_{54} & = & -dmp+dm^2n+cp+dq+2mn^2-cmn-m^3n-2np\nonumber\\
 & & {}-mq+m^2p+m^2q-m^3p+dr\nonumber\\
M_{53} & = & cq+2mnp-dnp-p^2+bp-nq+dm^2p-mr-dmq\nonumber\\
 & & {}-cmp+m^2q-m^3p+r\nonumber\\
M_{52} & = & bq-cmq-nr-dmr-dnq+cr+m^2r-m^3q+2mnq\nonumber\\
 & & {}-pq+dm^2q-m^3p+dr\nonumber\\
M_{51} & = & br-m^3r-dnr+dm^2r+2mnr-pr-cmr\nonumber\\
\end{eqnarray}  
These equations are then substituted into the equation,
$$\det(M)=\left|\matrix{
M_{11} & M_{12} & M_{13} & M_{14} & M_{15} \cr
M_{21} & M_{22} & M_{23} & M_{24} & M_{25} \cr
M_{31} & M_{32} & M_{33} & M_{34} & M_{35} \cr
M_{41} & M_{42} & M_{43} & M_{44} & M_{45} \cr
M_{51} & M_{52} & M_{53} & M_{54} & M_{55} \cr
}\right|=0$$
which generates a polynomial which will be reduced to Bring's equation(80).
Let this equation be,  
\begin{equation}
y^5+p_{4}y^4+p_{3}y^3+p_{2}y^2+Ay+B=0.
\end{equation}
Now take coefficients of the fourth, third, and second order terms in (67),
respectively, and set them to zero. Solving,
\begin{equation}
p_{4}=0
\end{equation}  
for $a$, one has
\begin{eqnarray}
a & = & 
\frac{1}{5}dm^3+\frac{1}{5}bm-\frac{3}{5}dmn-\frac{2}{5}n^2+\frac{4}{5}q-\frac{1}{5}cm^2+\nonumber\\
 & & {}+\frac{2}{5}cn-\frac{4}{5}mp+\frac{4}{5}m^2n+\frac{3}{5}dp-\frac{1}{5}m^4
\end{eqnarray}
Substituting $a$ and Bring's substitution for $b$ and $c$, i.e.
\begin{equation}
b=\alpha d+\xi
\end{equation}
\begin{equation}
c=d+\eta
\end{equation}
into(67), which becomes
\begin{equation}
y^5+Ay+B=0
\end{equation}
with the third and second order coefficients in (67), $p_{3}$ and $p_{2}$ set 
to
zero. $p_3$ is a quadratic in $d$, so that in may be written as,
\begin{equation}
p_{3}=p_{32}d^{2}+p_{31}d+p_{30}=0.
\end{equation}
Setting the coefficients, $p_{32}$, $p_{31}$ and $p_{30}$ each to zero, gives 
\begin{eqnarray}
p_{32} & = & 
(-\frac{2}{5}m^2+n)\alpha^2+(\frac{17}{5}m^2n-\frac{17}{5}mp+\frac{4}{5}m^3-2n^2-\frac{13}{5}nm-\frac{4}{5}m^4\nonumber\\
& &  
{}+3p+4q)\alpha+\frac{22}{5}m^2p+\frac{21}{5}mpn-\frac{19}{5}pn+\frac{12}{5}nm^4+5r-\frac{2}{5}m^6-\frac{18}{5}n^2m^2\nonumber\\
& &  
{}+2q+\frac{19}{5}n^2m-4m^3n+\frac{8}{5}m^2n+3qm^2-3mr-\frac{3}{5}p^2-3nq-5qm\nonumber\\
& &  
{}-2mp-\frac{2}{5}m^4-\frac{12}{5}m^3p+n^3+\frac{4}{5}m^5-\frac{3}{5}n^2\nonumber\\
& &   =0.
\end{eqnarray}
Since $p_{32}$ is quadratic in $\alpha$ we solve it for $\alpha$, and obtain,
\begin{eqnarray}
\alpha & = & \frac{1}{2}(-13nm-10n^2+4m^3+20q+17m^2n-4m^4+15p-17mp\nonumber\\
 & & {}+(-40qm^4+80qm^2+40m^3p+60np^2-15n^2m^4-190mpn-200nq\nonumber\\
 & & {}-15n^2m^2+400q^2+60n^3m^2-100n^2q-80mpn^2+200m^2r+260m^2nq\nonumber\\
 & & {}+225p^2-120m^3r+40m^5p+265m^2p^2-40qm^3-80m^4p-20qmn\nonumber\\
 & & {}+360m^2pn+30n^2m^3+600pq-510mp^2-120n^3m-680mpq\nonumber\\
 & & {}+300mnr-500nr+80pn^2-170m^3pn+60n^3))^\frac{1}{2}/(2m^2-5n).
\end{eqnarray}
So $\alpha$ is a number calculated directly from the coefficients of
the quintic.
Now the coefficient multiplying the linear term in $d$, $p_{31}$, is also linear in
both $\eta$ and $\xi$. Solving it for $\eta$ and substituting this into the zeroth
term in $d$, $p_{30}$, gives a quadratic equation in $\xi$, i.e. let
\begin{eqnarray}
p_{30} & = & \xi_{2}\xi^2+\xi_{1}\xi+\xi_{0}=0
\end{eqnarray}
and solve, giving,
\begin{eqnarray}
\xi & = & \frac{1}{2}(-\xi_{1}+(\xi_{1}^2-4\xi_{2}\xi_{0})^\frac{1}{2})/\xi_{2}.
\end{eqnarray}
Where the equations for $\xi_{0}$, $\xi_{1}$ and $\xi_{2}$, depend only on the
coefficients of the quintic, and are given in the pre-print${}^{27}$.
$p_{2}$ is cubic in $d$, i.e.
\begin{eqnarray}
p_{2} & = & d_{3}d^3+d_{2}d^2+d_{1}d+d_{0}
\end{eqnarray}
solving using Cardano's rule on Maple, gives
\begin{eqnarray}
d & = & \frac{1}{6}(36d_{1}d_{2}d_{3}-108d_{0}d_{3}^2-8d_{2}^3\nonumber\\
 & & {}+12\sqrt{3}(4d_{1}^3d_{3}-d_{1}^2d_{2}^2-18d_{1}d_{2}d_{3}d_{0}\nonumber\\ 
 & & 
{}+27d_{0}^2d_{3}^2+4d_{0}d_{2}^3d_{3})^\frac{1}{2})^\frac{1}{3}/d_{3}\nonumber\\
 & & 
{}-\frac{2}{3}(3d_{1}d_{3}-d_{2}^2)/(d_{3}(36d_{1}d_{2}d_{3}-108d_{0}d_{3}^2-8d_{2}^3\nonumber\\
 & & {}+12\sqrt{3}(4d_{1}^3d_{3}-d_{1}^2d_{2}^2-18d_{1}d_{2}d_{3}d_{0}\nonumber\\
 & & {}+27d_{0}^2d_{3}^2+4d_{0}d_{2}^3d_{3})^\frac{1}{2})^\frac{1}{3})\nonumber\\
 & & {}-\frac{1}{3}d_{2}/d_{3}.
\end{eqnarray}  
Where $d_{0}$, $d_{1}$, $d_{2}$, and $d_{3}$, are also given in the pre-print${}^{27}$.
Now (67) has been transformed to (72), where $A$ and $B$ depend only on 
the coefficients of the quintic(65). So
that equation(72),
with a linear transformation, becomes Bring's normal form,
\begin{eqnarray}
z^5-z-s & = & 0
\end{eqnarray}  
where,
\begin{eqnarray}
y & = & (-A)^\frac{1}{4}z
\end{eqnarray}
\begin{eqnarray}
s & = & - B/(-A)^{\frac{5}{4}}
\end{eqnarray}
We have used the fact that $p_{4}$ is linear in $a$ to get $p_{4}=0$. Then $b$, $c$ 
and $d$ were considered a point in space, on
the curve of intersection of a quadratic surface $p_{3}$, and a cubic surface, 
$p_{2}$. Giving $p_{3}=p_{2}=0$, as
required${}^{24, 26, 28}$.
\subsection*{B. The Solution to Bring's Normal Form}
Bring's normal form is solvable. Any polynomial that can be transformed to the form,
\begin{eqnarray}
z^n-az^m-b & = & 0
\end{eqnarray}
can be solved with the hypergeometric equation${}^{29}$. The solution is given 
by considering $z=z(s)$ and differentiating
Bring's equation (80) w.r.t. $s$ four times, then after making the substitution,
\begin{eqnarray}
t & = & \frac{5^5}{4^4}s^4
\end{eqnarray}
one obtains the fourth order Fuchian generalized hypergeometric differential 
equation,
\begin{eqnarray}
t^3(1-t)\frac{d^4z}{dt^4}+t^2(\frac{9}{2}-7t)\frac{d^3z}{dt^3}+t(\frac{51}{16}-\frac{411}{40}t)\frac{d^2z}{dt^2}+(\frac{3}{32}-\frac{183}{80}t)\frac{dz}{dt}+\frac{231}{160000}z=0\nonumber\\
\end{eqnarray}
whose solution is${}^{30-33}$,
\begin{eqnarray}
 z(s) = {}^{}_{\hphantom{0}4}{\rm
F}_{3}([\frac{-1}{20},\frac{3}{20},\frac{7}{20},\frac{11}{20}],[\frac{3}{4},\frac{1}{4},\frac{1}{2}],\frac{5^5}{4^4}s^4)
\end{eqnarray}
\subsection*{C. Reverse Tschirnhausian Transformation}
When undoing the Tschirnhausian transformation one must be careful to generate and 
keep all possible roots. All roots are calculated
below. Incorrect roots are eliminated by direct substitution. Firstly, calculate y, with (81). 
We now undo the Tschirnhausian transformation by substituting $d$, $c$, $b$, $a$ and 
$y$ into the quartic substitution, (64). The
resulting equation is solved using the quadratic method${}^{34}$. The quadratic method
introduces all the possible roots for the quartic. Let the general quartic,
\begin{equation}
x^4+a_{3}x^3+a_{2}x^2+a_{1}x+a_{0}=0
\end{equation}
have coefficients,
\begin{eqnarray}
a_{0} & = & a+y\nonumber\\
a_{1} & = &  b\nonumber\\
a_{2} & = &  c\nonumber\\
a_{3} & = &  d.\nonumber\\
\end{eqnarray}
There are twelve possible roots to the quartic, let them be contained in the array,
$x_{mn}$ with $m \in (1,2,3)$ and $n \in (1,2,3,4)$. Only four of these roots
satisfy
the quartic for a given set of coefficients $a_{0}$, $ a_{1}$, $a_{2}$ and $ a_{3}$.
Using nested loops to enter
the candidate solutions into both the general quintic (65) and the quartic(87), let the root that satisfies 
them 
both be 
$r_{1}$. This 
root is then factored out of the quintic(65), giving the following equations,
\begin{eqnarray}
a_{0} & = & q+r_{1}p+r_{1}^2n+mr_{1}^{3}+r_{1}^4\nonumber\\
a_{1} & = & p+r_{1}n+r_{1}^2m+r_{1}^3\nonumber\\
a_{2} & = & n+mr_{1}+r_{1}^2\nonumber\\
a_{3} & = & m+r_{1}\nonumber\\
\end{eqnarray}  
These equations are then substituted into the general quartic, which gives another
twelve roots, $x_{mn}$. This time four of these roots satisfy both the quartic and
quintic, let these roots be $r_{2}$, $r_{3}$, $r_{4}$ and $r_{5}$. They
were found the same way as $r_{1}$, using nested loops. The final
array has five non-zero elements, the five complex roots of the general quintic
equation(65).
\subsection*{D. Conclusion}
The five formulae for the roots of the general quintic equation are obtained,
without root ambiguity, in agreement with Gauss' `Fundamental Theorem of Algebra'  
when the polynomial is of degree five.
\section*{\uppercase\expandafter{\romannumeral6}. Determination of the Genus Two Hyperelliptic Periods for Arbitrary Parameters}
Now from the hyperelliptic function(49),
the periods can be calculated exactly for arbitrary branch places 
calculated in 
the previous 
section, this gives,
\begin{equation}
\omega'_{k,1}={u_{k}}^{c_{1},a_{1}}  = phef_{k}(c_{1},a_{1})+shef_{k}(c_{1},a_{1})
\end{equation}
where
\begin{eqnarray}
phef_{k}(c_{1},a_{1})= \frac{(A_{k,0}+A_{k,1}a_{1})}{i
\sqrt{(c-a_{1})(a_{2}-a_{1})(c_{2}-a_{1})}}{}\nonumber\\
{}\times
{F_{D}}^{(5)}(1,\frac{1}{2},\frac{1}{2},\frac{1}{2},\frac{1}{2},\frac{1}{2};2;\frac{c_{1}-a_{1}}{c-a_{1}},\frac{c_{1}-a_{1}}{a_{2}-a_{1}},\frac{c_{1}-a_{1}}{c_{2}-a_{1}},1,1)
\end{eqnarray}
\begin{eqnarray}
shef_{k}(c_{1},a_{1})= \frac{A_{k,1}(c_{1}-a_{1})}{2i
\sqrt{(c-a_{1})(a_{2}-a_{1})(c_{2}-a_{1})}}{}\nonumber\\
{}\times
{F_{D}}^{(5)}(2,\frac{1}{2},\frac{1}{2},\frac{1}{2},\frac{1}{2},\frac{1}{2};3;\frac{c_{1}-a_{1}}{c-a_{1}},\frac{c_{1}-a_{1}}{a_{2}-a_{1}},\frac{c_{1}-a_{1}}{c_{2}-a_{1}},1,1)
\end{eqnarray}
which becomes,
\begin{eqnarray}
{\omega'}_{k,1} = \frac{\pi (A_{k,0}+A_{k,1}a_{1})}{
2\sqrt{(c-a_{1})(a_{2}-a_{1})(c_{2}-a_{1})}}{F_{D}}^{(3)}(\frac{1}{2},\frac{1}{2},\frac{1}{2},\frac{1}{2};1;x_{1},x_{2},x_{3}){}\nonumber\\
{}+\frac{A_{k,1}(c_{1}-a_{1})\pi}{4
\sqrt{(c-a_{1})(a_{2}-a_{1})(c_{2}-a_{1})}}{F_{D}}^{(3)}(\frac{3}{2},\frac{1}{2},\frac{1}{2},\frac{1}{2};2;x_{1},x_{2},x_{3})
\end{eqnarray}
with,
\begin{equation}
x_{1}=\frac{c_{1}-a_{1}}{a_{2}-a_{1}};x_{2}=\frac{c_{1}-a_{1}}{c_{2}-a_{1}};x_{3}=\frac{c_{1}-a_{1}}{c-a_{1}}
\end{equation}
and similarly,
\begin{equation}
\omega_{k,2}=-{u_{k}}^{c,a_{2}}  =- phef_{k}(c,a_{2})-shef_{k}(c,a_{2})
\end{equation}  
\begin{eqnarray}
{\omega}_{k,2} = -\frac{(A_{k,0}+A_{k,1}a_{2})\pi}{2
\sqrt{(a_{1}-a_{2})(c_{2}-a_{2})(c_{1}-a_{2})}}{F_{D}}^{(3)}(\frac{1}{2},\frac{1}{2},\frac{1}{2},\frac{1}{2};1;x_{1},x_{2},x_{3}){}\nonumber\\
{}-\frac{\pi A_{k,1}(c-a_{2})}{4
\sqrt{(a_{1}-a_{2})(c_{2}-a_{2})(c_{1}-a_{2})}}{F_{D}}^{(3)}(\frac{3}{2},\frac{1}{2},\frac{1}{2},\frac{1}{2};2;x_{1},x_{2},x_{3})
\end{eqnarray}  
\begin{equation}
x_{1}=\frac{c-a_{2}}{a_{1}-a_{2}};x_{2}=\frac{c-a_{2}}{c_{2}-a_{2}};x_{3}=\frac{c-a_{2}}{c_{1}-a_{2}}
\end{equation}
and,
\begin{equation}
\omega'_{k,2}={u_{k}}^{c_{2},a_{2}}  = phef_{k}(c_{2},a_{2})+shef_{k}(c_{2},a_{2})
\end{equation}    
\begin{eqnarray}
{\omega'}_{k,2} = \frac{\pi (A_{k,0}+A_{k,1}a_{2})}{ 
2\sqrt{(a_{1}-a_{2})(c+a_{2})(c_{1}-a_{2})}}{F_{D}}^{(3)}(\frac{1}{2},\frac{1}{2},\frac{1}{2},\frac{1}{2};1;x_{1},x_{2},x_{3}){}\nonumber\\
{}+\frac{\pi A_{k,1}(c_{2}-a_{2})}{4
\sqrt{(a_{1}-a_{2})(c-a_{2})(c_{1}-a_{2})}}{F_{D}}^{(3)}(\frac{3}{2},\frac{1}{2},\frac{1}{2},\frac{1}{2};2;x_{1},x_{2},x_{3})
\end{eqnarray}
\begin{equation}
x_{1}=\frac{c_{2}-a_{2}}{a_{1}-a_{2}};x_{2}=\frac{c_{2}-a_{2}}{c-a_{2}};x_{3}=\frac{c_{2}-a_{2}}{c_{1}-a_{2}}
\end{equation}
and finally,
\begin{equation}
\omega_{k,1}={u_{k}}^{a,c_{1}}  = phef_{k}(a,c_{1})+shef_{k}(a,c_{1})
\end{equation}
\begin{equation}
\omega_{k,1} = {u_{k}}^{c_{2},a_{1}}+\omega_{k,2}
\end{equation}
where,
\begin{eqnarray}
u_{k}^{c_{2},a_{1}} = \frac{(A_{k,0}+A_{k,1}a_{1})\pi}{2\sqrt{(a_{2}-a_{1})(c_{1}-a_{1})(c-a_{1})}} 
{F_{D}}^{(3)}(\frac{1}{2},\frac{1}{2},\frac{1}{2},\frac{1}{2};1;x_{1},x_{2},x_{3}){}\nonumber\\
{}+\frac{A_{k,1}(c_{2}-a_{1})\pi}{4 \sqrt{(a_{2}-a_{1})(c_{1}-a_{1})(c-a_{1})}} 
{F_{D}}^{(3)}(\frac{3}{2},\frac{1}{2},\frac{1}{2},\frac{1}{2};2;x_{1},x_{2},x_{3})
\end{eqnarray}
\begin{equation}
x_{1}=\frac{c_{2}-a_{1}}{c-a_{1}};x_{2}=\frac{c_{2}-a_{1}}{c_{1}-a_{1}};x_{3}=\frac{c_{2}-a_{1}}{a_{2}-a_{1}}
\end{equation}
Riemann's moduli are given by,
\begin{eqnarray}
\tau_{1,1} & \equiv & \frac{\omega_{2,2} {\omega'}_{1,1}-\omega_{1,2} {\omega'}_{2,1}}{\Delta}\nonumber\\
\tau_{2,2} & \equiv & \frac{-\omega_{2,1} {\omega'}_{1,2}+\omega_{1,1} {\omega'}_{2,2}}{\Delta}
\end{eqnarray}
and by Riemann's condition(26),
\begin{eqnarray}
\tau_{1,2} & \equiv & \frac{\omega_{2,1} {\omega'}_{1,1}-\omega_{1,1} {\omega'}_{2,1}}{\Delta}\nonumber\\
 & \equiv & \frac{-\omega_{2,2} {\omega'}_{1,2}+\omega_{1,2} {\omega'}_{2,2}}{\Delta}\nonumber\\
 & \equiv & \tau_{2,1}
\end{eqnarray} 
\section*{\uppercase\expandafter{\romannumeral7}. The Solution to the Hyperelliptic $\phi$ Integral}
The solution to the hyperelliptic $\phi$ integral(7)is with $x_{1}=r_{g}/r$,
\begin{equation}
r=\frac{r_{g}{\theta_{12}}^2 \sqrt{a_{1}-a_{2}}}{c \sqrt{a_{1}-a_{2}} 
{\theta_{12}}^2-\sqrt{(a_{1}-c_{1})(a_{1}-c_{2})(a_{1}-c)} {\theta_{13}}^2}
\end{equation}
where $\theta_{13}$ and $\theta_{12}$ are given in (60) and (62). The 
hyperelliptic $\phi$ integral(10)may be written as
\begin{equation}
\frac{-\phi}{2}= \int_{a_{0}}^{x_{1}} \frac{A_{1,1}xdx}{y}\nonumber\\
=phef(A_{1,1};x_{1},a_{0})+shef(A_{1,1};x_{1},a_{0})
\end{equation}
\begin{eqnarray}
& &phef(A_{1,1};x_{1},a_{0}) =  B(a_{0})A_{1,1}\times {}\nonumber\\
&
&{}\sum_{(m)=0}^{+\infty}\frac{(1,m_{1}+m_{2}+m_{3}+m_{4}+m_{5})(\frac{1}{2},m_{1})(\frac{1}{2},m_{2})(\frac{1}{2},m_{3})(\frac{1}{2},m_{4})(\frac{1}{2},m_{5})}{(2,m_{1}+m_{2}+m_{3}+m_{4}+m_{5})m_{1}!m_{2}!m_{3}!m_{4}!m_{5}!}{}\nonumber\\
& & {} \times {x_{2}}^{m_{1}}{x_{3}}^{m_{2}}{x_{4}}^{m_{3}}{x_{5}}^{m_{4}}z^{m_{5}+1}
\end{eqnarray}
\begin{eqnarray}
& &shef(A_{1,1};x_{1},a_{0}) = \frac{1}{2}  B(r_{i_{1}}-a_{0})A_{1,1}\times 
{}\nonumber\\
&
&{}\sum_{(m)=0}^{+\infty}\frac{(2,m_{1}+m_{2}+m_{3}+m_{4}+m_{5})(\frac{1}{2},m_{1})(\frac{1}{2},m_{2})(\frac{1}{2},m_{3})(\frac{1}{2},m_{4})(\frac{1}{2},m_{5})}{(3,m_{1}+m_{2}+m_{3}+m_{4}+m_{5})m_{1}!m_{2}!m_{3}!m_{4}!m_{5}!}{}\nonumber\\
& & {} \times {x_{i_{2}}}^{m_{1}}{x_{3}}^{m_{2}}{x_{4}}^{m_{3}}{x_{5}}^{m_{4}}z^{m_{5}+2}
\end{eqnarray}
and
\begin{equation}
\frac{\phi'}{2}= \int_{a_{0}}^{x_{1}} \frac{(A_{2,0}+A_{2,1}x)dx}{y}
=phef(A_{2,0},A_{2,1};x_{1},a_{0})+shef(A_{2,1};x_{1},a_{0})
\end{equation}
\begin{eqnarray}
& &phef(A_{2,0},A_{2,1};x_{1},a_{0}) =  B\times (a_{0} A_{2,1}+A_{2,0}) {}\nonumber\\
&
&{}\sum_{(m)=0}^{+\infty}\frac{(1,m_{1}+m_{2}+m_{3}+m_{4}+m_{5})(\frac{1}{2},m_{1})(\frac{1}{2},m_{2})(\frac{1}{2},m_{3})(\frac{1}{2},m_{4})(\frac{1}{2},m_{5})}{(2,m_{1}+m_{2}+m_{3}+m_{4}+m_{5})m_{1}!m_{2}!m_{3}!m_{4}!m_{5}!}{}\nonumber\\
& & {} \times {x_{2}}^{m_{1}}{x_{3}}^{m_{2}}{x_{4}}^{m_{3}}{x_{5}}^{m_{4}}z^{m_{5}+1}
\end{eqnarray}
\begin{eqnarray}
& &shef(A_{2,1};x_{1},a_{0}) = \frac{1}{2} B A_{2,1}(r_{i_{1}}-a_{0})  \times {}\nonumber\\
&
&{}\sum_{(m)=0}^{+\infty}\frac{(2,m_{1}+m_{2}+m_{3}+m_{4}+m_{5})(\frac{1}{2},m_{1})(\frac{1}{2},m_{2})(\frac{1}{2},m_{3})(\frac{1}{2},m_{4})(\frac{1}{2},m_{5})}{(3,m_{1}+m_{2}+m_{3}+m_{4}+m_{5})m_{1}!m_{2}!m_{3}!m_{4}!m_{5}!}{}\nonumber\\
& & {} \times {x_{2}}^{m_{1}}{x_{3}}^{m_{2}}{x_{4}}^{m_{3}}{x_{5}}^{m_{4}}z^{m_{5}+2}
\end{eqnarray}
Abel's coefficients $A_{2,0}$ and $A_{2,1}$ depend on the cosmological constant and are determined by experimentation and observation. 
So with
\begin{equation}
x_{2}=\frac{a_{1}-a_{0}}{a_{2}-a_{0}};
x_{3}=\frac{a_{1}-a_{0}}{c-a_{0}};
x_{4}=\frac{a_{1}-a_{0}}{c_{1}-a_{0}};
x_{5}=\frac{a_{1}-a_{0}}{c_{2}-a_{0}}
\end{equation}  
where the branch places are related to the apsidal points by,
\begin{equation}
a_{1}=\frac{r_{g}}{r_{1}};a_{2}=\frac{r_{g}}{r_{2}};c=\frac{r_{g}}{r_{3}};c_{1}=\frac{r_{g}}{r_{4}};c_{2}=\frac{r_{g}}{r_{5}}
\end{equation}
and the variable place is given by,
\begin{equation}
z = \frac{x_{1}-a_{0}}{a_{1}-a_{0}}
\end{equation}
with
\begin{equation}
|x_{2}|<1;|x_{3}|<1;|x_{4}|<1;|x_{5}|<1;|z|<1
\end{equation}
so that the variables in the argument of the theta functions are,
\begin{eqnarray}
v_{1} & \equiv & \frac{-\omega_{1,2}\phi'+\omega_{2,2}\phi}{\Delta}\nonumber\\
v_{2} & \equiv & \frac{\omega_{1,1}\phi'-\omega_{2,1}\phi}{\Delta}
\end{eqnarray}
where the periods $\omega_{i,j}$ are given in section \uppercase\expandafter{\romannumeral6}. With this choice the equations of 
Forsyth and Baker are identical to 
these, in 
the case of $g=2$. The roots of the quintic $a_{1}$, $a_{2}$, $c_{1}$, $c_{2}$ and $c$ are given by setting the 
coefficients in(41) 
equal to
\begin{eqnarray}
\lambda_{0} & = & 4r = \frac{4\Lambda m^2 c^2 {r_{g}}^4}{3 L^2}\nonumber\\
\lambda_{1} & = & 4q = 0\nonumber\\
\lambda_{2} & = & 4p =  \frac{4}{3} \Lambda 
{r_{g}}^{2}-\frac{4m^2c^2{r_{g}}^2}{L^2}+\frac{4{{r_{g}}^2}E^2}{L^2 c^2}\nonumber\\
\lambda_{3} & = & 4n =  \frac{4m^2c^2{r_{g}}^2}{L^2}\nonumber\\
\lambda_{4} & = & 4m =  -4
\end{eqnarray}
then substituting into the solution to the quintic in section \uppercase\expandafter{\romannumeral5}.
\subsection*{A. Discussion}
The geodesics of a particle in SdS space are determined by three coordinates, the radial coordinate, $r$, and two angular 
coordinates 
$\phi$ and $\phi'$. $\phi$ is the actual physical coordinate, while $\phi'$ is an angular coordinate that comes from Abel's 
theorem. $r$ is a function of both $\phi$ and $\phi'$, by the quadrupoly periodic theta functions of two variables. Now $\phi$ and 
$\phi'$ depend on r, i.e. $\phi = \phi(r)$ and  $\phi' = \phi'(r)$ and they can be expressed as a linear combination of phef and shef. 
Thus $\phi = \phi(r) =-2phef(r)-2shef(r)$  is the closed form solution to the calculation of the geodesics in SdS 
space.
 \section*{\uppercase\expandafter{\romannumeral8}. The Reduction of the Hyperelliptic Functions to Elliptic Functions when 
$\Lambda=0$}
When the cosmological constant is set to zero  and the particles are subject to gravitation alone, two roots of the 
quintic(42) go to zero. Let these two roots be $c_{2}$ and $a_{2}$. Then ${\omega'}_{1,2}={\omega'}_{2,2}=0$. Then the 
following equations are obtained for the Riemann moduli,
\begin{equation}
\tau_{1,1} = (\omega_{2,2} {\omega'}_{1,1}-\omega_{1,2} {\omega'}_{2,1})/\Delta \neq 0
\end{equation}
\begin{equation}
\tau_{2,2} = (-\omega_{2,1} {\omega'}_{1,2}+\omega_{1,1} {\omega'}_{2,2})/\Delta = 0
\end{equation}
\begin{equation}
\tau_{1,2}  =   (\omega_{2,1} {\omega'}_{1,1}-\omega_{1,1} {\omega'}_{2,1})/\Delta\nonumber\\
            =  (-\omega_{2,2} {\omega'}_{1,2}+\omega_{1,2} {\omega'}_{2,2})/\Delta\nonumber\\
            =  0 
\end{equation}
Let Abel's coefficients be functions of the cosmological constant, i.e.
\begin{equation}
A_{1,1} \equiv A_{1,1}(\Lambda),
A_{2,0} \equiv A_{2,0}(\Lambda),
A_{2,1} \equiv A_{2,1}(\Lambda)
\end{equation}
Taking the limit as $\Lambda \rightarrow 0$ with (108), (111),(118), one has
\begin{equation}
v_{1} = \lim_{\Lambda \rightarrow 0} \frac{-\omega_{1,2} \phi' + \omega_{2,2} \phi}{\Delta}=0
\end{equation}
gives,
\begin{equation}
A_{2,0}(0)=0
\end{equation}
\begin{equation}
A_{2,1}(0) = \frac{-\omega_{2,2} }{\omega_{1,2}}
\end{equation}
and
\begin{equation}
v_{2} = \lim_{\Lambda \rightarrow 0} \frac{\omega_{1,1} \phi' - \omega_{2,1} \phi}{\Delta}=  \lim_{\Lambda \rightarrow 0} 
\frac{\phi}{\omega_{1,2}}
\end{equation}
or,
\begin{equation}
v_{2} =  \lim_{\Lambda \rightarrow 0} \frac{-2}{\omega_{1,2}} (phef(A_{1,1}(\Lambda);x,y)+shef(A_{1,1}(\Lambda);x,y))
\end{equation}
Letting $\lim_{\Lambda \rightarrow 0} A_{1,1}(\Lambda)= A_{1,1}(0)=\omega_{1,2}$ and using the identity,
\begin{eqnarray}zF_{D}^{(5)}(1,\frac{1}{2},\frac{1}{2},\frac{1}{2},\frac{1}{2},\frac{1}{2};2;x_{2},x,x,x_{3},z)\nonumber\\ 
-  \frac{1}{2}xz^{2}F_{D}^{(5)}(2,\frac{1}{2},\frac{1}{2},\frac{1}{2},\frac{1}{2},\frac{1}{2};3;x_{2},x,x,x_{3},z)\nonumber\\
= z F_{D}^{(3)}(1,\frac{1}{2},\frac{1}{2},\frac{1}{2};2;x_{2},x_{3},z)
\end{eqnarray}
to obtain the tripple hypergeometric function,
\begin{equation}
v_{2} = \frac{1}{2i} \frac{(x-y)}{\sqrt{(a_{1}-y)(c_{1}-y)(c-y)}} F_{D}^{(3)} 
(1,\frac{1}{2},\frac{1}{2},\frac{1}{2};2; 
\frac{a_{1}-y}{c_{1}-y},\frac{a_{1}-y}{c-y},\frac{x-y}{a_{1}-y})
\end{equation}
that reduces to the incomplete elliptical integral of the first kind${}^{35-36}$, when $y \rightarrow a_{1}$. So that $\phi$ 
becomes,
\begin{equation}
v_{2} = \phi = \frac{-2}{\sqrt{c-a_{1}}} F( sin^{-1} \sqrt{ \frac{x-a_{1}}{c_{1}-a_{1}}} , 
 \sqrt{\frac{c_{1}-a_{1}}{c-a_{1}}})
\end{equation}
 With these definitions,
\begin{equation}
\lim_{\Lambda \rightarrow 0} (\frac{\theta_{13}^{2}}{\theta_{12}^{2}})= \frac{\vartheta_{1}^{2}}{\vartheta_{4}^{2}}=\frac{1}{k} 
sn^{2}(v)
\end{equation}
where $sn(v)$ is Jacobi's elliptic function${}^{37}$. Now with the choice
\begin{equation}
c  =  \frac{r_{g}}{r_{1}},
c_{1}  =  \frac{r_{g}}{r_{2}},
a_{1}  =  \frac{r_{g}}{r_{3}},
k  =  \sqrt{\frac{r_{2} r_{1}-r_{3} r_{1}}{r_{2} r_{1}-r_{3} r_{2}}},
\end{equation}  
which gives the Schwarzschild black hole solution without the cosmological constant,
as a genus one elliptic function, which is the same as Ref. 7, it is
\begin{equation}
(\frac{1}{r}-\frac{1}{r_{1}})=(\frac{1}{r_{3}}-\frac{1}{r_{1}}) sn^2(v_{1})
\end{equation}
Lastly, the moduli become $1$ and
\begin{eqnarray}
\tau_{1,1} & = & \frac{{\omega'}_{2,1}}{\omega_{2,1}}\nonumber\\
& & {} =  \frac{\lim_{\Lambda \rightarrow 0} (phef_{2}(c_{1},a_{1})+shef_{2}(c_{1},a_{1}))}{\lim_{\Lambda \rightarrow 0} 
(u_{2}^{c_{2},a_{1}}+\omega_{2,2})}
\end{eqnarray}
Using the identities,
\begin{equation}
 F_{D}^{3}(\frac{1}{2},\frac{1}{2},\frac{1}{2},\frac{1}{2};1;x,x,z)- \frac{1}{2}  x 
F_{D}^{3}(\frac{3}{2},\frac{1}{2},\frac{1}{2},\frac{1}{2};2;x,x,z)=
{}^{}_{\hphantom{0}2}{\rm F}_{1}(\frac{1}{2},\frac{1}{2};1;z)
\end{equation}
\begin{eqnarray}
\pi F_{D}^{(3)}(\frac{1}{2},\frac{1}{2},\frac{1}{2},\frac{1}{2};1;1,x,y)
-\pi \frac{1}{2}F_{D}^{3}(\frac{3}{2},\frac{1}{2},\frac{1}{2},\frac{1}{2};2;1,x,y)
  = \frac{2}{\sqrt{x}} F(sin^{-1} \sqrt{x}, \sqrt{\frac{y}{x}})
\end{eqnarray}
where ${}^{}_{\hphantom{0}2}{\rm F}_{1}(\frac{1}{2},\frac{1}{2};1;z)$ is the Gauss function for the complete elliptical integral of 
the first kind and $F$ is the elliptic function for the incomplete elliptical integral of the first kind. Then the periods of section 
\uppercase\expandafter{\romannumeral6}, (90) to (104)reduce to elliptic functions, i.e.,
\begin{eqnarray}
{\omega'}_{k,1} & = &  A_{k,1}\frac{1}{\sqrt{c-a_{1}}} F_{D}^{(2)} ( \frac{1}{2}, \frac{1}{2}, \frac{1}{2}; \frac{3}{2};1, 
\frac{c_{1}-a_{1}}{c-a_{1}})\nonumber\\
& & {} = A_{k,1}\frac{1}{\sqrt{c-a_{1}}} \frac{\pi}{2} {}^{}_{\hphantom{0}2}{\rm 
F}_{1}(\frac{1}{2},\frac{1}{2};1;z)\nonumber\\ 
& & {} = A_{k,1}\frac{1}{\sqrt{c-a_{1}}} F(\frac{\pi}{2},\sqrt{\frac{c_{1}-a_{1}}{c-a_{1}}}),  
\end{eqnarray}
similarly,
\begin{equation}
u_{k}^{c_{2},a_{1}} \rightarrow u_{k}^{0,a_{1}} = A_{k,1} 
\frac{1}{\sqrt{c_{1}-a_{1}}} F(sin^{-1} \sqrt{\frac{a_{1}}{a_{1}-c}},\sqrt{\frac{a_{1}-c}{a_{1}-c_{1}} } )
\end{equation}
\begin{equation}
\omega_{k,2} = - A_{k,1} \frac{1}{\sqrt{c-c_{1}}} F 
(sin^{-1}(\sqrt{\frac{c}{c-a_{1}}}),\sqrt{\frac{c-a_{1}}{c-c_{1}}})
\end{equation}
\begin{equation}
\omega_{k,1} = u_{k}^{0,a_{1}}+\omega_{k,2}
\end{equation}
and lastly,
\begin{equation}
{\omega'}_{k,2} = 0
\end{equation}
It has been shown that the solution for the geodesics in SdS space in terms of 
multiple hypergeometric and hyperelliptic modular 
functions reduce to elliptic functions in the case of pure Schwarzschild space.
\section*{\uppercase\expandafter{\romannumeral9}. Conclusion}
The geodesics of particles in the combined Schwarzschild black hole and a non-zero cosmological constant, are 
given by 
ratios of quadrupoly periodic theta functions of two variables. The inverse functions are obtained as multiple 
hypergeometric functions of five variables, thus solving the Inversion Problem for 
hyperelliptic integrals of the first kind, i.e. 
genus two. It is a closed form solution since the branch places are 
found by solving the quintic equation for it's five roots. The five formulae for the five roots of the quintic were obtained, in 
agreement with Gauss' Fundamental Theorem of Algebra when the polynomial is of degree five. The closed 
form solution for the geodesics in SdS space was found 
to be $\phi=\phi(r)$, which is a linear combination of $phef(r)$ and $shef(r)$ which are convergent twelve slotted five variable 
hypergeometric functions.\\
\section*{Acknowledgments}
I would like to thank Kate Powell for proof reading and the 
referees for checking the work 
done in this manuscript.
\section*{}
{\small
\begin{flushleft}
${}^{1}$MacRobert, A.M., ``Sky $\&$ Telescope", June 2004, p.16; Riess, A.G., ApJ 607:665-687, 2004 June; Chae, K.H., ApJ 604:L71-L74 
June 
2004; Reiss, A.G., ApJ 116:1009-1038; Schmidt, B.P., ApJ 507:46-63; Perlmutter, S., ApJ 517:565-586.

${}^{2}$Drociuk, R.J. ``Cosmic Force'', http://xxx.lanl.gov/abs/gr-qc/0204023

${}^{3}$Witten, E., ``Anti-de Sitter Space, Thermal Phase Transition and confinement in 
gauge theories''.hepth/9803131, 
IASSNS-HEP-98/21,p.7(16 Mar 1998)

${}^{4}$Gibbons, G.W., Hawking, S.W., ``Cosmological event horizons, thermo\-dynamics, and 
particle creation'', Physical
Review D, Vol. 5, No. 10 pp. 2738-2756.(15 May 1977)

${}^{5}$Rindler, W., ``Essential Relativity'', Springer-Verlag, 1997.

${}^{6}$Filippenko, A.V., ``Einstein's Biggest Blunder? High-Redshift Supernovae and the 
Accelerating Universe'',
Publications of the Astronomical Society of the Pacific, Vol 113, pp 1441-1448, Dec. 2001.

${}^{7}$Barlett, J.H. ``Classical and Modern Mechanics'', The Univercity of Alabama 
Press, 1975 p.194.  

${}^{8}$Baker, H.F., ``Abelian Functions, Abel's theorem and allied theory of theta 
functions''. Cambridge Mathematical
Library. First Publication 1897 reissued in 1995

${}^{9}$Baker, H.F. ``Multiply Periodic Functions'', Cambridge University Press, 1907.

${}^{10}$Exton, H. ``Multiple Hypergeometric Functions and Applications'', John Wiley and 
Sons Inc. 1976. p.41,p.49,pp.
70-71.

${}^{11}$Landau, L.D., Lifshitz, E.M.,``Mechanics", Oxford, NYNY, Pergamon Press, 1976,$ 
3^{rd}$ Ed.

${}^{12}$Jacobi, von K.G.J.,``Ueber VIERFACH PERIODISCHEN FUNCTIONEN zweier Variabeln, 
auf 
die sich die Theorie der
Abel'shen Transcendenten stutzt".(Crelle's Journal fur reine und angewandte 
Mathematik Bd. 13,
1834.)(Herausgegeben von H. Weber aus dem Franzosischen ubersetzt von A. Witting.Leipzig verlag von wilhelm
engelmann 1895. Oftwald's Klaff der exakten Wiffenfchaften Nr. 64.

${}^{13}$Gauss, C.F. see his diary, also see Markshevich reference 15.

${}^{14}$ Abel, N.H.,``Oeures completes I, II", by L.Sylow and S. Lie, Grondal and Son, 
Christiania, new edition
1881.

${}^{15}$Markshevich, A.I., ``Introduction to the classical theory of Abelian functions", Translations of mathematical 
monographs; Providence, R.I., Amer. Math. Soc., 1992, v.96.p.1-2.

${}^{16}$Brioshi, F., ``Sur l' equation du sixieme degre", Acta mathematica, 12 Imprime le 
octobre 1888 (Milan).

${}^{17}$Bolza, O. `` Darstellung der rationalen ganzen Invarianten der Binarform sechsten 
Grades durch die Nullwerthe
der zugehorigen $\theta$-Functionen", Mathemat. Annalen, Leipzig, pp. 478-495. Gottingen, im Juni 1887.

${}^{18}$K\"onigsberger, Von Hern zu Greifswald, ``Ueber die Transformation der zweiten Abelschen
Functionen erster Ordnug.", Crelle Bd. 64 p.17-42. 1. April 1864. ``Ueber die Transformation des zweiten Grades f\"ur die Abelschen 
Functionen erster Ordnung" Crelle Bd. 67. p. 58-77.  Januar 1866.

${}^{19}$Hudson,R.W.H.T.``Kummer's Quartic Surface'', Cambridge Mathematical Library.

${}^{20}$Rosenhain, von Georg. ``Abhandlung uber die FUNCTIONEN ZWEIER VARIABLER mit vier 
Perioden, welche die
Inversen sind der ultra-elliptischen integrale erster Klasse"(Herausgegeben von H. 
Weber aus dem Franzosischen
ubersetzt von A. Witting.Leipzig verlag von wilhelm engelmann 1895. Oftwald's Klaffiker der exakten
Wiffenfchaften Nr. 65. This paper won the mathematics prize in Paris in the 1850's.

${}^{21}$Gopel, A.,``Theoriae transcendetium Abelianarum primi ordinis adumbratio 
levis'', Journal fur die reine und
angewandte Mathematik, Vol 35(1847).
  
${}^{22}$Forsyth, A.R. ``Introductory to the theory of Functions of Two Variables'', 
Cambridge at the University Press,
1914.

${}^{23}$Forsyth, A.R., ``Memoir on Theta Functions, particularly those of two variables, 
communicated by Prof. A.
Cayley'', Phil. Trans. 1881, p. 783.

${}^{24}$Bring, Erland Sam. B. cum D. Meletemata quae dam mathematica circa 
transformation aequationum  
algebraicarum, quae consent. Ampliss. Facult. Philos. in Regia Academia Carolina 
Praeside D. Erland Sam. Bring,
Hist. Profess-Reg. and Ord.publico Eruditorum Examini modeste subjicit Sven Gustaf Sommelius, Stipendiarius
Regius and Palmcreutzianus Lundensis. Die XIV. The main part of Bring's work is reproduced in Quar. Jour.  
Math., 6,
1864; Archiv. Math. Phys., 41, 1864, 105-112; Annali di Mat., 6, 1864, 33-42. There also is a re-publication of
his works in 12 volumes at the University of Lund.

${}^{25}$Harley, Rev. Robert,``A Contribution to the History of the Problem of the 
Reduction of the General Equation
of the Fifth Degree to a Trinomial Form." Quar. Jour. Math., 6, 1864, pp. 38-4

${}^{26}$Cayley, A. ``On Tschirnhaus's Transformation'', Phil. Trans. Roy. Soc. London, 
Vol. 151. pp.561-578. Also in
Collected Mathematical Papers, Cayley p. 275.

${}^{27}$Drociuk, R.J.  ``On the Complete Solution to the Most General Fifth
Degree Polynomial'', http://xxx.lanl.gov/abs/math.GM/0005026.

${}^{28}$Green, M.L. ``On the Analytic Solution of the Equation of Fifth Degree'', 
Composito Mathematica, Vol. 37, Fasc.
3, 1978. p. 233-241. sijthoff and Noordoff International Publishers-Alphen aan den. Netherlands.

${}^{29}$Weisstein, Eric W.`` CRC Concise Encyclopedia of Mathematics'', CRC Press, 1999. 
pp.
1497-1500.http://mathworld.wolfram.com/QuinticEquation.html

${}^{30}$Cockle, J.``On Transcendental and Algebraic Solution-Supplementary Paper". Phil. 
Mag. Vol. 13, pp.135-139.

${}^{31}$Harley, R.``On the theory of the Transcendental Solution of Algebraic 
Equations". 
Quart. Journal of Pure and
Applied Math. Vol. 5 p.337. 1862.

${}^{32}$Cayley, A. ``Note on a Differential Equation", Memoirs of the Literary and 
Philosophical Society of
Manchester, vol. II(1865),pp. 111-114. Read February 18, 1862.

${}^{33}$Slater, L.J. ``Generalized Hypergeometric Functions'', Cambridge University 
Press, 1966, pp. 42-44.

${}^{34}$Drociuk, R.J. ``On the Root Ambiguity in the complete solution to the Most 
General Fifth Degree Polynomial''
http://xxx.lanl.gov/abs/math.GM/0207258

${}^{35}$Exton,H.,''Handbook of Hypergeometric 
Integrals,Theory,Applications,Tables,Computer Programs'', John Wiley and 
Sons,1978.

${}^{36}$Carlson, B.C.``Some series and bounds for complete elliptical integrals." J. Math. and Physics. 40,(1964), 125-134.

${}^{37}$Gradshteyn, I.S.,Ryzhik,I.M.``Table of Integrals, Series, and Products" $5^{th}$ Ed., Academic Press, 1994.
\end{flushleft}}
\end{document}